\documentclass[twocolumn,aps,floatfix,prb]{revtex4}
\usepackage{graphics, epsfig ,color} 
\usepackage{amsmath} 
\usepackage{amsfonts}

\begin{document}
\date{\today}
\title{Improving Students' Lab Practices:  the Performance Grade}
\author{G.L. Lippi$^{1,2}$}
\address{$^1$ Universit\'e de Nice Sophia Antipolis, Institut Non Lin\'eaire de Nice\\
$^2$ CNRS, UMR 7335\\
1361 Route des Lucioles, F-06560 Valbonne, France}
\email{Gian-Luca.Lippi@inln.cnrs.fr}

\begin{abstract}
Instilling good laboratory working attitudes in students is a difficult but very important task, especially in the first level courses.  The introduction of a grade, based on the {\it observation} of work practices during laboratory sessions, can be strongly beneficial towards the acquisition of positive skills covering not only the technical aspects, but also the acquisition of both independence and team work.  Explicit suggestions are given for basing the grade on specific observations and a quantitative analysis is performed to guarantee that the higher intrinsic volatility of the Performance Grade does not affect the final laboratory grade.
\end{abstract}

\maketitle

\section{Introduction} 

Bad habits are better avoided than unlearnt:  this dictum applies to laboratory practices as well.  Instructors are often confronted with the problem of how to convince and/or encourage students to take a positive attitude towards their practical experience and maximize their learning, rather than concentrating their attention on the short-term reward:  the grade.

Given that laboratory work is for the most part a relatively new and foreign experience for a large number of students beginning their studies in Physics (and more generally, Science), on the one hand one has the privilege of building upon a (fairly clean) slate but, on the other hand, one also encounters the difficult task of convincing the students to take up good habits.  Although building on rather pristine ground may sound like an ideal situation, this often turns out not to be the case, since the average student does not have a good idea of what experimental work truly means.  Thus, at times it may be rather difficult to make oneself understood.  Compounded with the pressure of obtaining good scores, the communication hurdle can become frustrating for the instructor and damaging for the student.

For all the explanations about what is expected, in an average class the instructor finds herself/himself confronted with the (hopefully only) occasional case of the student who at the end of the course still has not got it right.  And this is not necessarily for lack of trying from either side.  Indeed, the feedback to the student arrives at best at the following session (typically one week later) and since laboratory courses -- in particular at first level -- tend to have a large attendance, the instructors may change from one session to the next.  Thus, the transmission of information becomes problematic and it is all too natural that some students may take a ``wait and see" attitude.

The standardized system of education, as it is universally employed at least in western countries, uses a grading system not only as a form of evaluation, but also of feedback.  In spite of all the possible criticism which can be raised to its use, it is certain that students are well attuned to the grade, since they have beeen exposed to it through all their schooling.  Thus, the idea of using a form of grading to give feedback to students and signal whether they are on the right track seems to be a reasonable one.  After all, this is what is done in theoretical courses where homework is graded weekly.  

The practical objection which is often raised is how to do this in a fair and reliable way.  I am going to present a framework in which such a grade, which I am going to call the {\it performance} or {\it participation grade}~\cite{per-part}, can be satisfactorily used for the benefit of students without impinging on the fairness of the overall laboratory course evaluation.  I have introduced this method of encouraging students' participation and guide their laboratory work about ten years ago in undergraduate labs in our Physics program and have worked with numerous collegues -- among others several Teaching Assistants and Temporary Instructors.  The experience has been positive and has allowed to more effectively guide a much larger percentage of students towards developing a positive working attitude in Experimental Physics.

\section{The Performance or Participation Grade}

The Participation Grade is an evaluation given at the end of each laboratory session by the Laboratory Instructor(s) of the quality of the work performed by each individual student in the course of the session.  Its main features are summarized in Table~\ref{sum}.  
The main goals to be assessed are the student's attitude towards learning and her/his active involvement in experimental work which should progressively lead to the acquisition of the automatic reflex of continuously questioning both procedures and results, rather than passively accepting any obtained results or all the instructions received.  This in turn develops critical thinking and builds a strong foundation for the development of the basic skill expected of any physics graduate:  the ability to analyze new situations and find solutions to new problems.  
Of course, the degree of success has to be reasonably assessed in accordance with the course level (i.e., introductory vs. advanced lab) but clearly requiring this from students is beneficial at all levels.  Another important point in the evaluation is the kind of interaction that each individual student has both with the Instructor and with her/his Peers:  a student always asking for help, for the {\it right answer}, for {\it what needs to be done now} is clearly on the wrong track.  In the same way, a student passively following a leader -- e.g. the lab partner -- is not standing up to the expected level.  However, as stressed in Section~\ref{use}, a student leading the group by neglecting her/his partner(s) is also not performing up to standards, since Laboratory work is supposed to be a group effort.  This setting offers therefore a unique opportunity to learn about (scientific) interaction and collaboration.

\begin{table}
\caption{
Synoptic definition of the characteristics of the {\it Performance} or {\it Participation Grade}.  The first category (top box) highlights the goals each student has to strive for, the second box the nature of the grade (individual, even though the work is collective), the last three categories present the strenghts and weakness of this kind of grade.
}
\label{sum}
\begin{tabular}{|| c | l ||}\hline\hline
& Attitude \\
& Questioning \\
Goals & Critical thinking \\
& Interaction and group work \\
& Consulting sources (advanced labs)~\cite{cons} \\ \hline\hline
Nature & Individual grade \\ \hline\hline
Effectiveness & Quantitative assessment \\ \hline
Immediacy & Feedback at the end of session \\ \hline
Weakness & Subjectivity and reliability (fluctuations) \\ \hline\hline
\end{tabular}
\end{table}

By its nature, the Participation Grade is individual, even though laboratory work is of a collective nature.  The reason for this apparent dichotomy is that through collective work
{\it each student} has to conquer the outlined goals, thus the feedback provided by the Performance grade must be tailored to the individual.  Students may be at first surprised to be individually evaluated for group work, but once the scope is clearly explained they will appreciate its importance and will react positively.  This grade operates therefore as a positive feedback mechanism for each individual within the ensemble.  Transmitted to each student at the end of the session, together with clear explanations and specific recommendations (in oral or written form), this assessment gives an immediate return on the less-tangible goals (i.e., those not related to getting the correct results from the experiment), by providing individual and specific directions to acquire good laboratory practices.

\begin{table}
\caption{
Global (but not necessarily complete) summary of Instructor actions suggested for assessing the student's Performance Grade.  The left column classifies the actions as indirect (observation without interaction with students) and direct (resulting from direct interactions).  Both kinds of actions should be performed by every Instructor during each lab session for the students of which s/he has charge.  The right column gives a non-exhaustive list of specific points to be observed for arriving to the grade assessment.  The symbols ``+" and ``-" in front of various points in the analyical discussion (r.h.s.) stand for positive and negative attitudes which should either be rewarded or sanctioned.  As mentioned in the text, these are guidelines rather than rigid sets of checkpoints, and the grade should be given in broad categories (thus mitigating the burden on the Instructor).  
}
\label{ind}
\begin{tabular}{|| c | l ||}\hline\hline
Action & Description\\ \hline
& Observe the interactions among students:\\
Indirect & any member leading and/or dominating the group? \\ 
& any member left out? \\ \hline
& Observe the degree of involvement:  \\
& is everyone participating in the work?  \\
Indirect & to what degree?\\
& \hspace{2.5cm}
\begin{tabular}{l l}
+ & taking turns \\
+ & discussing the work \\
- & just writing the results\\
- & sitting back and waiting\\
\end{tabular}
\\ \hline
Direct & Kind of questions asked of the Instructor:\\
& \hspace{2.5cm}
\begin{tabular}{l l}
+ & trying to understand? \\
+ & actively looking for problem? \\
- & asking for a ready solution? \\
- & prying out the right answer?\\
\end{tabular}
\\
& who is participating in the exchange? \\ \hline
Direct & Questions/challenges posed by the Instructor:\\&  {\bf a.} globally, to the group \\& {\bf b.} directly, to an individual student\\
& \hspace{0.6cm}
\begin{tabular}{l l}
Technical & Instrument operation \\
& Uncertainties and tolerances \\
& Pitfalls \\
Fundamental & Understanding of processes\\
Comprehension & Intuitive description\\
& Further developments \\
& Problems\\
\end{tabular}
\\
\hline\hline
\end{tabular}
\end{table}

The Instructor's assessment should be based on two broad categories of indicators~\cite{unpub}, summarized in Table~\ref{ind}.  The actions which are listed in the table are a non-exhaustive list of the points which can and should be examined by the Instructor.  The breadth of the task, although mitigated by the grading scheme suggested below (Section~\ref{constr}), immediately indicates the difficulty in giving a quantitatively reliable grade.  Fluctuations must be expected, due to the sometimes large number of students and groups (and even different kinds of experiments a single Instructor has to follow), to students' behaviour (some groups will be more demanding than others), to more or less frequent requests for help by some groups, to the need for troubleshooting problems with instrumentation, etc.  Thus, the task appears to be daunting and may discourage anybody from using this kind of grading scheme, were it not for the fact that it is possible to set up a weighting system which renders it at the same time effective for the student and practically {\it uninfluential} for the final grade, thus lifting any possible anxiety the Instructors may feel about using it.  In the following we will prove how it is possible to set up the scheme in such a way that the reliability of the course grade is not affected by the use of the Performance Grade.

Achieving this goal will also ensure that any criticism related to the higher intrinsic volatility of the Performance Grade, due to the perception-based assessment given by the Instructor rather than on grading a written document, will lose any value and the full benefits of this grade may be reaped.

\section{Grade construction}\label{constr}

In the context of the grading scheme, and in order to simplify the discussion, we are going to consider the total grade assigned to each student for a Laboratory Course to be composed of an Ordinary Grade, $G_o$, and a Participation (or Performance) Grade, $G_p$:
\begin{eqnarray}
\label{gradecomp}
G_t = \sum_{i=1}^M w_o G_{o,i} + \sum_{j=1}^N w_p G_{p,j} \, ,
\end{eqnarray}
where $w_o$ and $w_p$ are the weights which are assigned to the ordinary and participation grades, respectively.  For the sake of generality, and in order to exploit one of its advantages for improving the total grade reliability, we assume that the total number of {\it ordinary grades} be $M \ne N$, with $N$ number of {\it participation grades}.  The meaningful inequality is $M < N$, which is equivalent to saying that the lab is organized in such a way as to give more than one participation grade per lab session (cf. discussion in Section~\ref{impact}).

Throughout the paper we consider the total grade $G_t$ normalized to 1 (maximum value) and use percentages to represent actual grades.  This choice has the advantage of rendering the discussion independent of the grading system, which changes from one country to another (and may not even be uniform even among different Universities of a same country).  The Reader will make the necessary adjustments to convert the values given into the units of her/his University for specific use.

The simplification that we introduce in grouping under $G_o$ all the aspects of ordinary grading does not restrict the generality of the discussion, but simply serves to set out $G_p$ from the rest.  If the Ordinary Grade is composed, for instance, of graded Lab Reports and a final exam, or of grades on the Labbook and separate Lab Reports, or whatever other combination, then it suffices to decompose $G_o$ further (an example of more complex grade composition has been discussed in the context of grading accuracy~\cite{Lippi2014}).

Fluctuations (even strong ones) in the Performance Grade have to be expected.  It will be therefore convenient for the Lab Supervisor to give the team of Lab Instructors some common rules for the Participation Grade, specifying semi-quantitative criteria based on Table~\ref{sum} (or on variations of the criteria exposed there).  
Predetermining a small number of possible grades {\bf not} based on the usual grading system but oriented towards more constructive feedback for the student is also quite helpful in establishing the Performance Grade.
A possible set may be:  {\it Excellent, Good, Acceptable and Insufficient}.  These broad categories -- which can be cast in different form (e.g., letters in a numerical grading system) -- have the double advantage of:  1. partially diverting the students' attention from the detailed grade and 2. allowing, when needed, for grade renormalization when several Instructors participate in the lab supervision.  For instance, a Lab Supervisor who notices that particular Instructors tend to give excessively high, or low, evaluations compared to their collegues, may weight the Participation Grades issued by the different collegues before transforming them into the usual grading scale.

In the following section we discuss in detail the influence of the (admittedly larger) fluctuations in $G_p$ on the accuracy of the global grade $G_t$ and what constraints can be applied to ensure that these fluctuations do not impinge on standard deviation, $\sigma_{G_t}$, of the grade, i.e., on its reliability.  One additional precaution is to plan the grading scheme with the largest reasonable number of values for $G_p$, since in the averaging process one can expect to obtain an estimate for $G_p$ whose reliability grows with the number of events.  Changing Instructors, whenever possible, also ensures better averaging.  Indeed, as discussed in the following section, it may be useful to have, if possible, two Instructors assign -- independently of each other -- a $G_p$ to each student for every lab session.

\section{Impact of the Performance grade on the final grade and on its accuracy}\label{impact}

An assessment of the impact of the participation grade and its uncertainty on the reliability of the global grade can be obtained using an {\it a priori} analysis~\cite{Lippi2014}.  
Using the proposed grade composition, eq.~(\ref{gradecomp}) with $w_o + w_p = 1$, we further assume that each grade be attributed with respective uncertainty $\sigma_o$ and $\sigma_p$.  To simplify the discussion, we assume each uncertainty to be homogeneous over the ensemble of grades of the same kind (e.g., all Performance Grades are affected by the same $\sigma_p$).  However, this constraint does not restrict the validity our analysis and, if needed, can easily be relaxed~\cite{Lippi2014}.  Notice that by using the statistical description of the uncertainties, we implicitely attribute a gaussian nature to the problem~\cite{Taylor1982}.
%
%

Our discussion is based on the size of the uncertainty on the participation grade $\sigma_p$ relative to $\sigma_o$.  These uncertainties can only be estimated {\it a priori}~\cite{Lippi2014} and in order to fix ideas~\cite{Lippi2014}, we will set $2 \sigma_o = 0.03$.  This choice amounts to saying that the ordinary grade $G_o$ is guaranteed within $\pm$3\% with probability 95\%, the usual error margin set in risk analysis (as discussed~\cite{Lippi2014}, Section 4) -- nearly 100\% (more precisely 99.7\%) would require~\cite{Taylor1982} $\pm 3 \sigma$.  The following analysis, however, is entirely general and holds for any value of $\sigma_o$.  This example is just a way of having a reasonable reference for interpreting the discussion which follows.


The {\it a priori} estimate for $\sigma_p$ in principle requires direct knowledge of the conditions under which the supervision is conducted.  However, as we show below, it is possible to determine conditions ensuring that the variability of the Participation Grade does not reflect onto that of the total grade $G_t$, whose uncertainty remains solely determined by $\sigma_o$.

Following~\cite{Lippi2014}, we easily arrive to the expression for the combined uncertainty on the global lab grade
\begin{eqnarray}
\sigma_t & = & \sqrt{\frac{w_o^2 \sigma_o^2}{M} + \frac{w_p^2 \sigma_p^2}{N}} \, , \\
& = & \frac{w_o \sigma_o}{\sqrt{M}} \sqrt{1 + \frac{M}{N} \left( \frac{w_p}{w_o} \right)^2 \left( \frac{\sigma_p}{\sigma_o} \right)^2} \, , \\
\label{unc-corr}
& \approx & \frac{w_o \sigma_o}{\sqrt{M}} \left[ 1 + \frac{1}{2} \frac{M}{N} \left( \frac{w_p}{w_o} \right)^2 \left( \frac{\sigma_p}{\sigma_o} \right)^2 \right] \, ,
\end{eqnarray}
where in writing eq.~(\ref{unc-corr}) we have implicitely assumed the relative weight coefficient of the participation grade, $\frac{w_p}{w_o}$, to be small so as to use a first-order approximation to the square root.

In the present context, the approximation is not a simple mathematical assumption to simplify the calculation:  its nature reflects the {\it requirement} that the participation grade should not impinge on the accuracy of the grade and represents the {\it constraint} that we want to impose to ensure fairness.  Thus, we require the second term in the bracket, eq.~(\ref{unc-corr}), to be small.  For the sake of generality, we fix its value to be 
\begin{eqnarray}
\frac{1}{2} \frac{M}{N} \left( \frac{w_p}{w_o} \right)^2 \left( \frac{\sigma_p}{\sigma_o} \right)^2 & \leq & u \, ,
\end{eqnarray}
where the value of $u$, small, can be later chosen.  This allows us to find a constraint on the maximum value, $\sigma_{p,max}$, of the uncertainty on the participation grade as a function of the other parameters:
\begin{eqnarray}
\label{spmax}
\left(\frac{\sigma_{p,max}}{\sigma_o}\right)^2 & = & 2 u \frac{N}{M} \left(\frac{w_p}{w_o}\right)^2 \, .
\end{eqnarray}

\begin{figure}[ht!]
\hglue -.5cm
\includegraphics[width=\linewidth,clip=true]{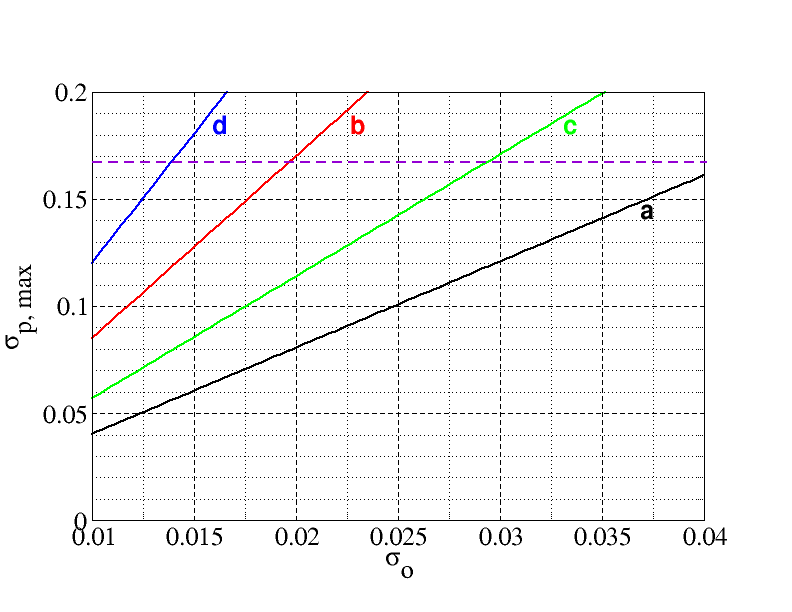}
\vglue -0.7cm
\caption{
Maximum value of the uncertainty on the participation grade to satisfy the {\it fairness} criterion.  $u = 0.1$ (cf. text).  The various curves, from eq.~(\ref{spmax}), are traced for the following set of parameters:  (a) $N=M$, $w_0 = 0.9$, $w_p=0.1$; (b) $N=M$, $w_0 = 0.95$, $w_p=0.05$; (c) $N=2 \times M$, $w_0 = 0.9$, $w_p=0.1$; (d) $N=2 \times M$, $w_0 = 0.95$, $w_p=0.05$.  The dashed, horizontal line represents the value of $\sigma_{p,max}$ for which the interval $\pm 3 \sigma_{p,max} = 1$ (i.e., the uncertainty is as large as the full grading scale).  The choice of $\sigma_o$ discussed at the beginning of this section corresponds to finding the intersections between a vertical straight line at $\sigma_0 = 0.015$ and the estimates for $\sigma_{p,max}$.
}
\label{sigmap-max}
\end{figure}

\underline{\it Fairness criterion} (choice of $u$):  to ensure that the impact of the Performance Grade on the {\it reliability} of the total lab grade be negligible it suffices to choose $u=0.1$, since it renders the value of $\sigma_p$ one order of magnitude smaller~\cite{upperval} than that of $\sigma_o$.  Since uncertainties are rounded to the first digit~\cite{Taylor1982}, the composition of the uncertainties~\cite{Lippi2014} ensures that the contribution coming from $\sigma_p$ -- thus appearing only on the second digit -- will always be negligible.  This way, one does not have to worry about the larger variability of the participation grade which, given the numerous constraints (number of students, changing instructors, etc.), may be more liable to stronger fluctuations.

Fig.~\ref{sigmap-max} shows the maximum value permitted for the uncertainty on the participation grade $\sigma_{p,max}$, as a function of the uncertainty of the ordinary grade, $\sigma_o$, for different values of participation grades $N$ ($\frac{N}{M}$ representing the number of participation grades per lab session) and different weights $w_p$ ($w_o = 1- w_p$).  The horizontal, dashed line, represents the maximum possible value for $\sigma_{p,max}$, i.e., the value for which the interval $\pm 3 \sigma_{p,max}$ covers the whole range of possible grades (i.e., from 0\% to 100\%).  This corresponds to the limit where a Performance Grade randomly attributed (with a gaussian distribution), rather than given with some pedagogical criterion, does not not have an impact on the accuracy of the overall grade.  Obviously, this limit is absurd since even a poorly organized and inefficient Participation Grade average will do better than that, but it provides a well-defined border beyond which the curves lose meaning.  We therefore take the horizontal dashed line as the cutoff for all the curves.

The lowest curve (a, black online) represents the most difficult situation in which to fulfill the fairness criterion, since the Performance Grades is given a (relatively) large weight (10\%) and the number of Performance Grades equals the number of Ordinary Grades ($N=M$).  For an Ordinary Grading Uncertainty equalling 1\% (i.e., guaranteeing that the grade is absolutely accurate at  $\pm 6$\%, quite a stringent constraint!), one can only afford value of $\sigma_{p,max} = 0.04$ (i.e., total uncertainty $\pm$ 12\%), which may be rather difficult to satisfy.  However, for a more reasonable uncertainty in the ordinary grade (e.g. $\sigma_o = 0.02$, which guarantees a total interval of 8\% with 95\% confidence~\cite{Taylor1982} -- i.e., $\pm 2 \sigma_o$), $\sigma_{p,max}$ already grows to 0.08.  This is a comfortable value, since it implies that globally, the uncertainty on the true value grade is not affected by the participation grade unless its overall ($3 \sigma$) uncertainty grows beyond $\pm$ 24\%.  Larger values of $\sigma_o$ are traced for completeness, but are not expected to play a serious role, since a worst-case error in evaluation of $\pm$24\%, averaged over all lab sessions and lab instructors, is already quite large on a 100\% scale! 

When the number of participation grades is doubled (curve (c), green online), the conditions are immediately relaxed, due to the $\sqrt{\frac{N}{M}}$ contribution in eq.~(\ref{spmax}).  At $\sigma_o = 0.02$ (cf. above discussion), $\sigma_p$ may take values as large as .115, thus bringing the $3 \sigma_{p,max}$ interval to 35\% (thus the $\pm 3 \sigma$ interval to 70\%!).  The other two curves refer to smaller weights for $w_p = 0.05$, which at $\sigma_o = 0.02$ already renders the uncertainty on the participation grade entirely irrelevant (its maximum value is the whole grade interval) even when the number of participation grades equals that of ordinary grades (curve (b), red online).  Eq.~(\ref{spmax}) shows that the dependence on $u$ and on $\sqrt{\frac{N}{M}}$ is weak (due to the square root) and therefore the cases covered by the curves of Fig. \ref{sigmap-max} are fairly complete; the changes in values coming from different choices of $u$, $M$ or $N$ will not substantially affect the expected $\sigma_{p,max}$.  

Thus, we conclude that the {\it fairness} constraint can be quite easily satisfied by reasonable choices of the relative weights for the grades, with the chosen values of $u$ and $\frac{N}{M}$.

{\bf Choice of value for $w_p$}:  I personally prefer choosing the larger $w_p = 0.1$ because of its psychological value.  The beginning student, to whom this system is primarily addressed, is not aware of the negligible impact that the participation grade has on the accuracy of the final outcome (although I would encourage teaching advanced students the use of this analysis~\cite{Lippi2014}).  Thus s/he is more likely to focus on the straight contribution of the participation grade $G_p$ to the total and a 10\% contribution represents more of an incentive towards good performance.  

Nuances exist, due to the differences coming from the grading system (letters, percentages, numbers $\ldots$) and from local traditions, and these can be best appreciated by the experienced Lab Supervisor, who can suitably adapt the ideas exposed here to the local context and wisely choose the value to attribute to $w_p$.  The aim is to maximize the student's motivation while maintaning a high degree of reliability of the final grade.

\section{Using the participation grade}\label{use}

The grade should encourage good work and immediately warn students against bad attitudes, such as waiting passively, trying to get the right answer by simple trial and error -- e.g., by prodding the instructor to see what the correct answer may be --, letting the partner(s) do the work and sitting back $\dots$.  At the same time, it should also be used to instill the correct working practices for team collaboration.  In a group where strong level disparities appear, one needs to push the weaker student not to sit back and profit from the partner(s), but also signal the stronger student(s) to include the weaker ones and help them progress.  This is a very important lesson which anyone can learn and which goes well beyond the framework of the laboratory course.  Leaving aside moral and societal considerations -- which in spite of their importance cannot necessarily be included in lab instruction -- it will be clear to any student that the ability of working in teams will be paramount to their future success, independently on their career choices.  Thus, a negative evaluation -- low participation grade -- can be correctly given to student(s) who may have personally excelled during lab time but have entirely disregarded and perhaps have marginalized one of more of their team members.

In order to make full use of this grade, it is important that either the individual Instructor or the Lab Supervisor (depending on class sizes and organization) reserve time -- preferably at the end of each session --, to discuss with the students their Performance Grade, what are the reasons which have lead to the choice that has been made and which aspects of the work and attitude need improvement.

\section{A few special cases}

One of the most difficult cases to handle in lab is the student who is exclusively focussed on getting the {\it right} result, to {\it guarantee} for her/himself a good grade.  Most instructors -- typically Graduate Students or Temporary/Junior Faculty members -- are told by the Lab Supervisor in charge of the whole course not to give the answers but let the students work it out for themselves.  However, the student entirely focussed on the grade will not easily let go.  The participation  grade, given at the end of each session and explained to the students, is a good way of steering grade-motivated students.  Receiving a bad grade (a 0\% should not be discouraged since it can be psychologically \underline{very effective}, while bearing little impact on the final grade) right at the end of the session accompanied by an explanation of the reasons for the bad grade can work wonders for instilling the {\it right motivation}, particularly into those students who are grade-driven and who wouldn't otherwise be deterred by other arguments!

Another kind of course where we have successfully used the Participation Grade is in advanced labs for students in their last year before the degree.  There, students are required to start taking some initiative and perform (partially) independent work, on the basis of suggestions given in laboratory writeups.   Not all students, even though they may have well mastered the more technical aspects of previous lab courses, readily turn to independent work; paradoxically, some of those best at performing under guidance may have more difficulties in gaining independence.  Thus, the Participation Grade serves the purpose of giving quick quantitative feedback all along to more effectively guide and encourage the development of independence in lab.

\section{Conclusions}

The Performance Grade offers the potential for guidance to students in their attitude towards learning laboratory techniques, independence, critical thinking but also acquiring group-working practices and skills.  The evaluation of the student performance during lab sessions is in itself a complex task, open to different and varying influences which may change with Instructor, working conditions and day-to-day events.  As such, fluctuations in the evaluation are unavoidable and may lead to worries about the volatility of this grade and its negative influence on the final lab grade.  I have shown that by appropriately choosing procedures and weight coefficients, one can limit this influence below leves which ensure that the uncertainty on the final grade remains entirely unaffected by the larger intrinsic fluctuations of the Participation Grade.  

As a final observation, the Performance Grade will influence the lab grade of each individual student, thus {\it lifting the degeneracy} in the group, even though this modification in itself will not be very large (due to the small weight $w_p$).  This should be regarded as a positive contribution, since it is important to include in the evaluation of laboratory practices the aspects which have been listed in Tables~\ref{sum} and \ref{ind}.

\acknowledgements

The author acknowledges the collaboration of all students who have taken the two semesters of undergraduate physics laboratory course at the Universit\'e de Nice-Sophia Antipolis for the past ten years and all the collegues -- young and less young , but too numerous to be individually mentioned --, who have served as Instructors.


\begin{thebibliography}{10}
\bibitem{cons} In an advanced lab it is often appropriate and recommended to have students consult additional documentation, such as web ressources, instrumentation manuals, textbooks,  or scientific articles.  The independence, initiative, degree of involvement and performance in accomplishing this tasks during lab work (if sufficient time is planned in the session) can be an integral part of the participation grade assessment.
\bibitem{upperval} The statement holds strictly for the values of $\sigma_o$ which we realistically consider (cf. Fig.~\ref{sigmap-max}).  If the significant digit in $\sigma_o$ were allowed to take values larger or equal to 5, then the rounding procedure would increase the uncertainty on $\sigma_t$.  However, this case is not considered since it would require unreasonably high uncertainties on $\sigma_o$.
\end{thebibliography}
\end{document}